\documentclass[prl,twocolumn,superscriptaddress,showpacs,floatfix]{revtex4-1}

\usepackage[final]{graphicx}
\usepackage{amsfonts,amssymb,amsmath}

\usepackage{color}
\definecolor{red}{rgb}{1.0,0.0,0.0}
\definecolor{blue}{rgb}{0.0,0.0,1.0}
\definecolor{green}{rgb}{0.0,1.0,0.0}

\newcommand{\etal}{\emph{et al.}}

\newcommand{\mrm}[1]{\ensuremath{\mathrm{#1}}}

\newcommand{\hc}[1]{\ensuremath{#1^\dagger}}

\newcommand{\ti}{\tilde}

\newcommand{\al}[1]{\begin{align} #1 \end{align}}

\newcommand{\bT}[1]{\ensuremath{\left\{ #1 \right\}}}

\newcommand{\bS}[1]{\ensuremath{\left[ #1 \right]}}

\newcommand{\abs}[1]{\lvert #1 \rvert}

\newcommand{\pd}[1]{\partial_{#1}}

\newcommand{\hbm}{\hbar^{-1}}

\newcommand{\om}{\omega}

\newcommand{\sig}{\sigma}

\begin{document}

\title{
Microscopic theory of indistinguishable single-photon emission from a quantum dot coupled to a cavity: The role of non-Markovian phonon-induced decoherence
}

\author{P. Kaer}\email{per.kaer@gmail.com}
\affiliation{DTU Fotonik, Department of Photonics Engineering, Technical University of
Denmark, Building 345W, 2800 Kgs. Lyngby, Denmark}
\author{P. Lodahl}
\affiliation{Niels Bohr Institute, University of Copenhagen, Blegdamsvej 17, 2100 Copenhagen, Denmark}
\author{A.-P. Jauho}
\affiliation{CNG, DTU Nanotech, Department of Micro- and Nanotechnology Engineering,
Technical University of Denmark, Building 345E, 2800 Kgs. Lyngby, Denmark}
\author{J. Mork}
\affiliation{DTU Fotonik, Department of Photonics Engineering, Technical University of
Denmark, Building 345W, 2800 Kgs. Lyngby, Denmark}

\date{\today}

\begin{abstract}
We study the fundamental limit on single-photon indistinguishability imposed by decoherence due to phonon interactions in semiconductor quantum dot-cavity QED systems.
Employing an exact diagonalization approach we find large differences compared to standard methods.
An important finding is that short-time non-Markovian effects limit the maximal attainable indistinguishability.
The results are explained using a polariton picture that yields valuable insight into the phonon-induced dephasing dynamics.
\end{abstract}


\maketitle

\begin{figure}[h]
 \centering
 \includegraphics[width=\columnwidth]{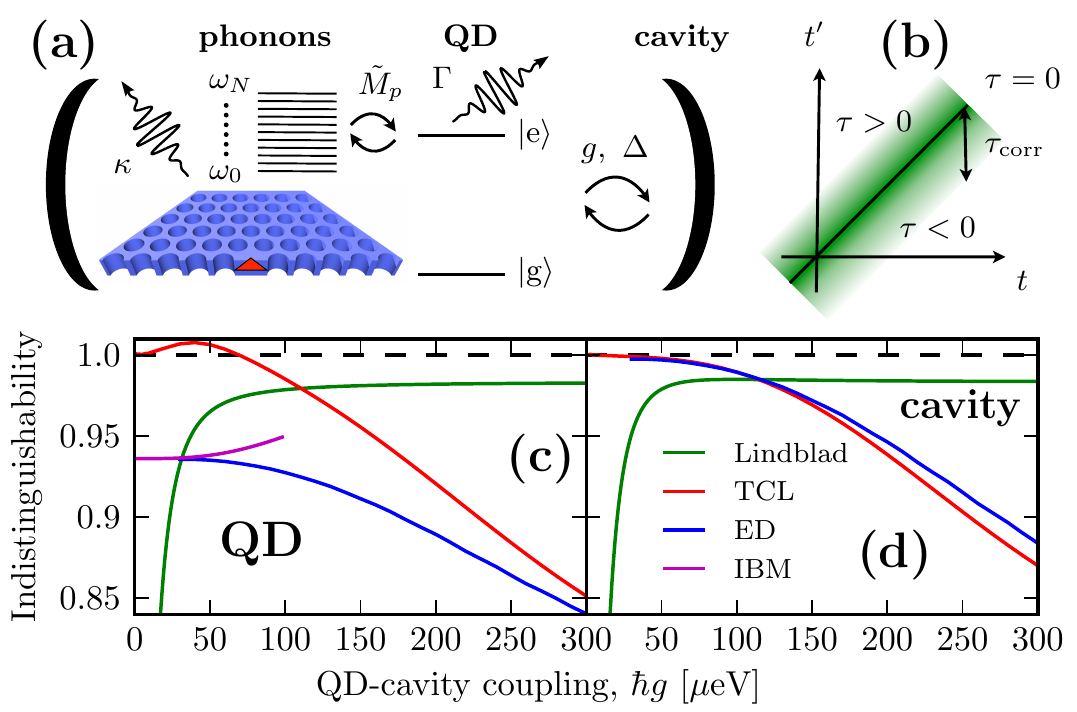}
 \caption{\textbf{(a)} Schematic of a coupled QD-cavity system interacting with
 longitudinal acoustical phonons and a photonic crystal cavity with an embedded QD. \textbf{(b)} Two-time plane with the time diagonal,
 $\tau=0$, and phonon reservoir correlation time $\tau_\mathrm{corr}$. The shaded region
 shows the extent of the short-time regime. \textbf{(c)}, \textbf{(d)} Calculated indistinguishability as a function of the QD-cavity coupling strength for light emitted from the QD and the cavity.
Parameters: $\Gamma=0.5~$ns$^{-1}$, $\hbar\kappa=125~\mu$eV, $\hbar\Delta=27.78~\mu$eV, and $\hbar\gamma=0.85~\mu$eV.}
\label{fig:multi_fig_system_twotime_ID_g}
\end{figure}

The study of the coherence properties of single photons emitted
from semiconductor cavity QED (cQED) systems is
important for applications in quantum information
technology \cite{Knill2001} and provides insight into the fundamental
decoherence effects induced by the environment.
For all-solid-state cQED systems, such as a quantum dot (QD) embedded in
a photonic crystal cavity \cite{Calic2011} [Fig. \ref{fig:multi_fig_system_twotime_ID_g}(a)]
or a micropillar cavity \cite{Madsen2011},
the main decoherence mechanism
at low temperatures
is the electron-phonon interaction \cite{Borri2001,Besombes2001,Muljarov2004},
as many recent studies show \cite{Milde2008,Winger2009,Calic2011,Hohenester2009c,Kaer2010,Madsen,Majumdar2011,Roy2011a,Ulrich2011}.

Decoherence limits the degree of indistinguishability of single photons emitted from cQED systems [Fig. \ref{fig:multi_fig_system_twotime_ID_g}(a)], thus diminishing their applicability for scalable linear optical quantum computing \cite{Knill2001}, where an all-solid-state single-photon source is a key element.
Furthermore, recent experimental results \cite{Winger2009,Calic2011,Hohenester2009c,Madsen,Majumdar2011,Ulrich2011} necessitate a departure from the well understood paradigms of atomic cQED, since the strong interaction with reservoirs in the solid state calls for new basic models and physical interpretations.
A better understanding of phonon-induced decoherence
thus leads to insight into the fundamental physics of nanostructured solids, and
can help ushering novel quantum technological devices.
However,
thus far only little attention has been given to the influence of phonon interactions on the indistinguishability.
Only few experiments have been reported \cite{Santori2002,Varoutsis2005,Madsen2011} and previous theoretical studies have employed a Markovian pure dephasing approximation \cite{Bylander2003,Kiraz2004,Ben2005,Yao-Yi2005,Troiani2006,Sun2009,Cancellieri2009,
Pathak2010} or phenomenological descriptions of finite-memory dephasing processes \cite{Santori2009}, none of them treating the phonon interaction microscopically while taking into account the cavity.

In this Letter we show that the
non-Markovian nature of the phonon reservoir has a large
effect on single-photon indistinguishability:
short-time virtual
processes occurring on
time scales
much shorter than a typical ``dephasing time'',
must be considered.
Also, it is essential to treat the phonon interaction
microscopically and on equal footing with the electron-photon
interaction.
The analysis is based on an
exact diagonalization (ED) technique, retaining the inherent non-Markovian nature of the phonon interaction to all
orders in the phonon coupling.
Our findings are contrasted to standard approximate
approaches for including phonon interactions
\cite{Naesby2008,Yamaguchi2008,Auffeves2009,Hohenester2010,Roy2011a}, namely second
order expansions and phenomenological pure dephasing descriptions.
Figures \ref{fig:multi_fig_system_twotime_ID_g}(c) and (d) show such a comparison. The deviations between the approximate and the ED results are significant, demonstrating that memory and back-action effects in the reservoir cannot be neglected, as
in Markovian approaches.

To calculate the indistinguishability, we model the celebrated Hong-Ou-Mandel experiment \cite{Hong1987,Santori2002} where two-time correlation functions for the photon operator $a$ need to be considered \cite{Kiraz2004}, $\langle a^\dag(t')a(t)\rangle=\langle a^\dag(t+\tau)a(t)\rangle$.
These are defined in the plane spanned by $t$ and $t'$ [Fig. \ref{fig:multi_fig_system_twotime_ID_g}(b)], whereas one-time correlation functions, $\langle a^\dag(t)a(t)\rangle$, reside on the time diagonal ($\tau=0$).
Consequently, one-time functions only experience short-time non-Markovian phonon effects within  the phonon reservoir correlation time, $\tau_\mathrm{corr}$ [Fig. \ref{fig:multi_fig_system_twotime_ID_g}(b)], after the initial excitation.
In contrast, the two-time function implies that a photon is removed, $a(t)$, at \emph{each} instant $t$ and added again, $a^\dag(t+\tau)$, an instant $\tau$ later.
This results in a ``continuous excitation'' of short-time transients, illustrated as the band surrounding the time-diagonal in Fig. \ref{fig:multi_fig_system_twotime_ID_g}(b).
Short-time non-Markovian effects thus play an important role throughout the entire lifetime of the excitation, strongly affecting physical quantities derived from two-time functions, such as the indistinguishability which we will demonstrate.

\emph{Theory.} -- To model a QD coupled to a cavity interacting  with longitudinal acoustical phonons, we employ the Jaynes-Cummings model
 including the electron-phonon interaction \cite{Wilson-Rae2002,
Milde2008, Hohenester2009c, Kaer2010} [Fig.
\ref{fig:multi_fig_system_twotime_ID_g}(a)].
We follow Hohenester \cite{Hohenester2007} and employ a set of effective phonon modes.
Expanding the QD-cavity system in the basis  $\bT{|1\rangle=|\mrm e,n=0\rangle,|2\rangle=|\mrm g,n= 1\rangle,|3\rangle=|\mrm g,n=0\rangle}$, where $n$ is the cavity photon number, the total Hamiltonian
becomes
\al{\label{eq:H_full}
H=H_\mrm{JC} + \sum_{p} \ti M_{p}(\ti{b}^\dagger_{p} + \ti b_{p}) \sig_{11} + \sum_{p}
\hbar \om_{p} \ti{b}^\dagger_{p}\ti b_{p},
}
where $\sig_{ij}=|i\rangle\langle j|$. The second term describes the phonon
interaction with the QD, $p$ denotes the effective phonon modes with bosonic
operators $\ti{b}_{p}$ and $\ti{b}^\dagger_{p}$, and $\ti M_p$ is the effective phonon
matrix element \footnote{$\ti M_p=\bS{4\pi\Delta k_{p}k^{2}_{p}V/(2\pi)^{3}}^{1/2}M_{k_{p}}$, where $k_{p}$ is the radial
wavevector for mode $p$, $\Delta k_{p}=k_{p+1}-k_{p}$. $M_{k} = \bS{\hbar k/(2\rho_\mrm{d}c_\mrm{s}V)}^{1/2}\bS{D_
\mrm{e}-D_\mrm{g}}\exp\bS{-\frac 14(kl)^2}$ is the matrix element for bulk
phonons, assuming equal wavefunctions for the excited and ground state in the QD. We use GaAs parameters: $\rho_\mrm d=5370~\mrm{kg}\mrm{m}^{-3}$, $c_\mrm
s=5110~\mrm{m}\mrm{s}^{-1}$, $D_\mrm e=-14.6~\mrm{eV}$, $D_\mrm g = -4.8~
\mrm{eV}$, and $l=5~\mrm{nm}$}. The last term is the free phonon Hamiltonian with $
\om_{p}$ denoting the frequency of mode $p$ \footnote{We assume a linear dispersion
relation given as $\om_{p}=c_\mrm{s}k_{p}$, where $c_\mrm{s}$ is the speed of sound.}.
The Jaynes-Cummings model is
\al{\label{eq:H_JC}
H_\mrm{JC}=\hbar \Delta \sig_{11} + \hbar g (\sig_{12} + \sig_{21}),
}
where $\Delta=\om_\mrm{eg}-\om_\mrm{cav}$ is the QD-cavity detuning with
$\om_\mrm{eg}$ and $\om_\mrm{cav}$ being the QD and cavity transition frequencies,
respectively, and $g$ is the QD-cavity coupling strength.

\begin{figure}[h]
 \centering
 \includegraphics[width=\columnwidth]{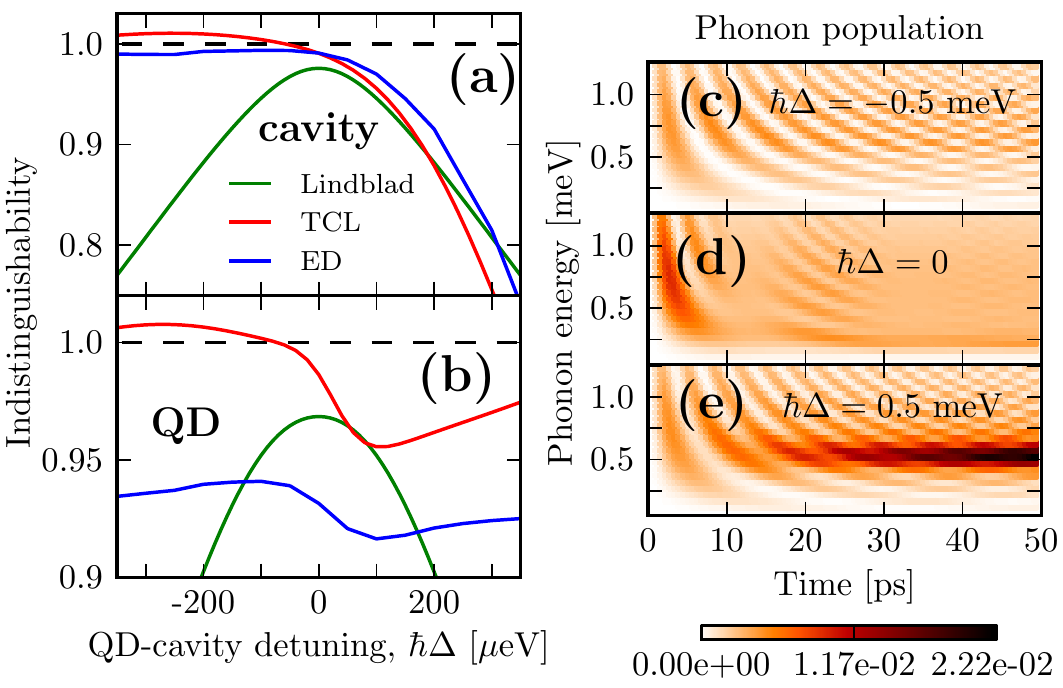}
 \caption{\textbf{(a)}, \textbf{(b)} Indistinguishability as a function of QD-cavity detuning
 for QD and cavity emission. \textbf{(c)}, \textbf{(d)}, and \textbf{(e)} Phonon population
 distribution function for different QD-cavity detuning. For $\Delta=0$ the population has been  scaled by a factor of 2. Parameters: $\hbar\Gamma=1~\mu$eV, $\hbar \kappa=100~\mu$eV, $\hbar g=100~\mu$eV, and $\hbar\gamma=1.1~\mu$eV.}
\label{fig:ID_vs_detun}
\end{figure}

The system dynamics is obtained by employing the reduced density matrix formalism
including Lindblad decay terms \cite{Carmichael1999,Breuer2002}. The Master equation for
the density matrix is
\al{\label{eq:rho_EOM}
\pd t \rho(t) = -i\hbm\bS{H, \rho(t)}+S_\mrm{L}\rho(t),
}
where the Lindblad terms are $S_\mrm{L}\rho(t)=(L_{\kappa}\bT{\sig_{32}}+L_{\Gamma}\bT{\sig_{31}})\rho(t)$.
The rate $\kappa$ describes the escape of cavity photons, related to the $Q$-factor as $\kappa=Q/\om_\mrm{cav}$, and the rate $\Gamma$ describes the decay of the QD in the absence of the cavity.
The Lindblad operator is $L_{\eta}\bT{O}\rho(t)=-\frac\eta2\bS{\hc OO\rho(t)+\rho(t)\hc OO-2O\rho(t)\hc O}$.
Importantly, the electron-phonon interaction is here included in the unitary part of the Master equation and not via approximate scattering terms.
Thus, the electron-phonon interaction is treated on equal footing with the electron-photon interaction, ensuring a rigorous inclusion of all non-Markovian phonon effects.
To solve Eq. (\ref{eq:rho_EOM}), we expand the phonon operators in a multi-phonon Fock state basis \footnote{The typical number of included phonon modes was 30 to 50, with up to 2 phonon excitations in each mode, resulting in 200 to 500 phonon states. Sampling the 3 phonon excitation space, we estimate an error of at most 0.1 percent in Fig. \ref{fig:multi_fig_system_twotime_ID_g}(c) and (d) and at most 1 percent in Fig. \ref{fig:ID_vs_detun}(a) and (b).},
and propagate the equations numerically, providing an exact diagonalization of the coupled QD-cavity-phonon system.
In the limit of $g \rightarrow 0$, our model becomes the exactly solvable independent boson model (IBM) \cite{Mahan1993,Nazir2009b,Goan2010,Goan2011a}.
The IBM can thus extrapolate the ED results to this limit, where the ED becomes cumbersome due to long numerical integration times.

We compare our simulations with two
standard approaches. The first treats the
electron-phonon interaction to second order using the time-convolutionless method (TCL)
\cite{Breuer1999,Breuer2002,Kaer2012c}.
Here the density operator is
$\bar\rho(t)=\mrm{Tr}_\mrm{phon}\bS{\rho(t)}$, where the phonons are
traced out and hence treated as a thermal reservoir.
Within this approximation, the Master equation becomes
\al{\label{eq:rhobar_EOM}
\pd t \bar\rho(t) = -i\hbm\bS{H_\mrm{JC}, \bar\rho(t)}+S_\mrm{TCL}(t)\bar\rho(t)+
S_\mrm{L}\bar\rho(t),
}
where the effects of the phonons are contained in $S_\mrm{TCL}(t)$ \cite{Kaer2010,Kaer2012c}.
The second approach is a Markovian Lindblad description of the pure dephasing processes,
equivalent to the TCL for a
 memory-less phonon reservoir.
The consequence is the replacement of $S_\mrm{TCL}(t)$ with the Lindblad operator
$L_{2\gamma}\bT{\sig_{11}}$, where $\gamma$ is the pure dephasing rate \cite{Naesby2008,Yamaguchi2008,Auffeves2009}.
The rate $\gamma$ will be chosen to provide a reasonable fit to the ED.

To calculate two-time functions we invoke the Quantum Regression Theorem (QRT) \cite{Carmichael1999}, which does not imply any approximations in describing non-Markovian phonon effects in the ED approach.
Applying the QRT to the TCL density matrix in Eq. (\ref{eq:rhobar_EOM}), however, requires more subtle considerations.
The TCL results in time-dependent scattering rates \cite{Breuer1999,Kaer2010} and thus the substitution $\gamma_\mrm{TCL}(t)\rightarrow\gamma_\mrm{TCL}(\tau)$ might be expected to include non-Markovian effects in the QRT.
This is however not the case, as recently shown by Goan \etal~\cite{Goan2010,Goan2011a}.
However, in the long-time limit, $S_\mrm{TCL}(t\rightarrow\infty)$, the QRT becomes a consistent approximation.

The indistinguishability of the emitted photons is quantified as the normalized number of coincidence events at the HOM output detectors and is calculated as \cite{Kiraz2004,Pathak2010}
\al{
I = \frac{\int_{0}^{\infty}dt\int_{0}^{\infty}d\tau\abs{\langle \hc A(t+\tau)A(t)\rangle}^{2}}
{\int_{0}^{\infty}dt\int_{0}^{\infty}d\tau\langle \hc A(t+\tau)A(t+\tau)\rangle
\langle \hc A(t)A(t)\rangle},
}
where $A$ is either the photon operator
$a=|n=0\rangle\langle n=1|=\sig_{32}$ for light emitted from the cavity, or the QD
operator $\sig_{-}=|g\rangle\langle e|=\sig_{31}$ for light emitted from the QD.
We assume a 1-photon basis and second-order contributions are therefore absent.

For all simulations, except when varying the detuning, a QD-cavity detuning equal
to the polaron shift \cite{Kaer2010} has been used, $\hbar\Delta=\hbar\Delta_\mrm{pol}\approx~27.78~\mu\mrm{eV}$, corresponding to an effective QD-cavity detuning close to zero.
To investigate the fundamental limits on indistinguishability set by the vacuum phonon bath we set $T=0~$K.
We emphasize that neither our model nor approach are limited to zero temperature, however, finite temperatures significantly increase the computational effort.
The initial condition is the QD in the excited state, with both the photon and phonon fields in their ground states, corresponding to the experimental situation of excitation of the system with a short optical pulse, usually employed in measurements of the indistinguishability.
We neglect effects such as timing jitter and nearby fluctuating charges, as these depend on the excitation mechanism and can be avoided.

\emph{Dependence of QD-cavity coupling strength.} -- Figures
\ref{fig:multi_fig_system_twotime_ID_g}(c) and (d) show the indistinguishability for light
emitted from the cavity and QD as a function of the QD-cavity coupling strength $g$.
The ED and IBM results \footnote{$I_\mrm{IBM,QD}=\Gamma_\mrm{eff} \int_0^\infty d\tau \exp (-\Gamma_\mrm{eff} \tau-2\mrm{Re}[\varphi(0)-\varphi(\tau)])$, where $\mrm{Re}[\varphi(\tau)]=\sum_{p} |\tilde M_p/(\hbar \omega_p)|^2\cos (\omega_p \tau)$, and with $\gamma=0$ in the definition of $\Gamma_\mathrm{eff}$ as phonon dephasing is included microscopically.} differ quantitatively and qualitatively from both the TCL and Lindblad results.
For small but increasing $g$
the indistinguishability remains constant in the ED and IBM approaches for the QD, which is also expected for the cavity, where the IBM does not apply.
The TCL may predict an indistinguishability above unity [Fig. \ref{fig:multi_fig_system_twotime_ID_g}(c)], which is unphysical and a well-known issue associated with this method \footnote{See pp. 127-131 in \cite{Breuer2002}}.
Further increasing
the QD-cavity coupling,
$g$,
the indistinguishability
decreases for the ED and TCL, whereas the Lindblad theory predicts saturation.
The surprisingly large deviations between the ED and common approaches are important, especially for applications with strict requirements on the indistinguishability, e.g., in implementations of linear quantum computing protocols \cite{Knill2001}.

The Lindblad theory can qualitatively be interpreted using
\cite{Bylander2003,Sun2009}
$I=\Gamma_\mathrm{eff}/(\Gamma_\mathrm{eff}+2\gamma)$,
where $\Gamma_\mathrm{eff}$ is the effective decay rate of the QD. For large detuning or
small QD-cavity coupling, compared to the loss rates, one obtains \cite{Auffeves2009}: $\Gamma_\mathrm{eff}=\Gamma+2g^{2}\gamma_\mathrm{tot}/(\Delta^{2}+
\gamma^{2}_\mathrm{tot}),~\gamma_\mathrm{tot}=1/2(\kappa+\Gamma+2\gamma)$.
The expression predicts an initial increase in indistinguishability with coupling strength due to the Purcell effect, until it saturates for $\hbar g>100~\mu$eV as the strong coupling regime is entered.
The results from the ED and TCL clearly cannot be explained using this model.

\begin{figure}[h]
 \centering
 \includegraphics[width=\columnwidth]{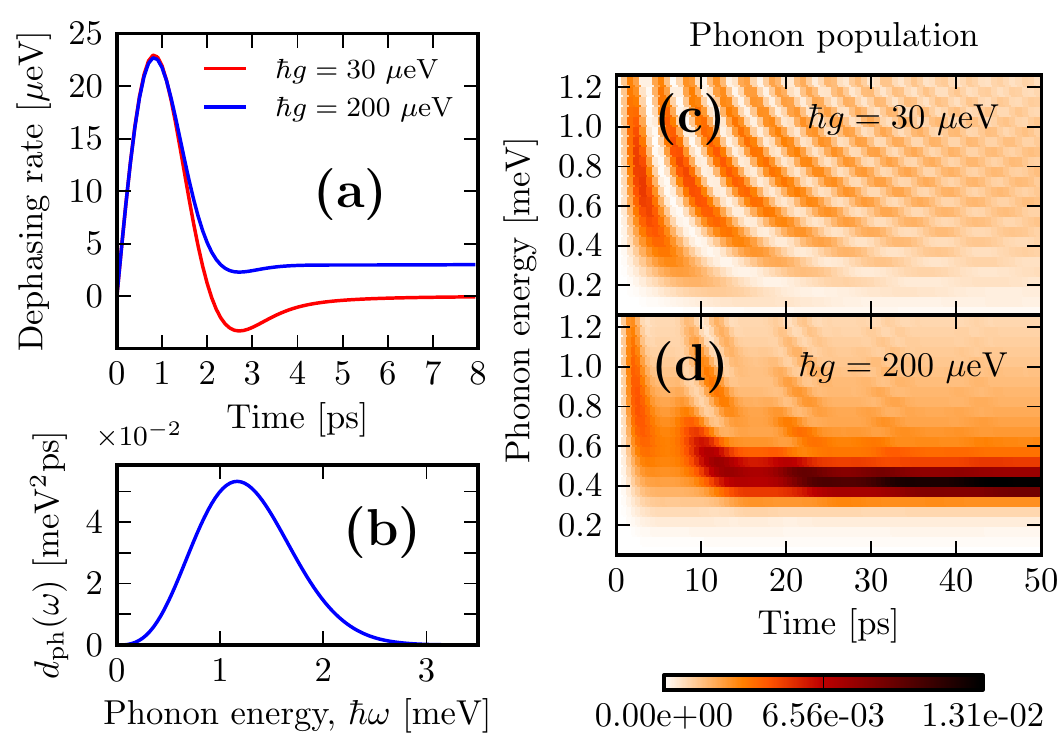}
 \caption{\textbf{(a)} Time-dependent dephasing rate from the TCL. \textbf{(b)} Effective
 phonon density at zero temperature. \textbf{(c)} Phonon population distribution function
 for $\hbar g=30~\mu$eV, $\Gamma=0.5~$ns$^{-1}$, $\hbar\kappa=125~\mu$eV, and
 $\hbar\Delta=27.78~\mu$eV. \textbf{(d)} As \textbf{(c)} with $\hbar g=200~
 \mu$eV.}
\label{fig:multi_fig_g_explain}
\end{figure}
It is instructive to write the Hamiltonian, Eq. (\ref{eq:H_full}), in terms of polariton (dressed) states, which diagonalize the Jaynes-Cummings Hamiltonian, Eq. (\ref{eq:H_JC}).
This results in terms like $\sig_\mathrm{lu}\sum_{p}\ti M_{p}(\ti{b}^\dagger_{p}+\ti b_{p})$, which cause phonon-mediated transitions between the upper (u) and lower (l) polariton
branches.
These are separated by an energy $\om_\mathrm{u}-\om_\mathrm{l}=\sqrt{4g^{2}+\Delta^{2}}$, and hence this energy is expected to play an important role in the physical
interpretation.

Figure \ref{fig:multi_fig_g_explain}(a) shows examples of time-dependent
dephasing rates calculated within the TCL, for two values of the coupling strength, $g$.
Figure \ref{fig:multi_fig_g_explain}(b) shows the corresponding effective phonon density, defined as $(T=0)$
$d_\mathrm{ph}(\om)=\pi\sum_{p}\abs{\ti{M}_{p}}^{2}\delta(\om-\om_{p})$.
For both values of $g$, the dephasing rate attains large values within the first 3 ps, after which it settles to a smaller positive non-zero value.
The initial temporal variations of the rate are directly related to a sampling of the entire effective phonon density through virtual processes, which are allowed at short times due to the energy-time uncertainty relation.

In the long-time limit, phonon-induced decoherence reflects real phonon-mediated transitions, corresponding to the effective phonon density being sampled at specific energies.
This explains why the long-time value of the dephasing rate is much larger for $\hbar g=200~\mu$eV compared to $\hbar g=30~\mu$eV.
The phonon density is thus sampled, respectively, at the energies given by the polariton transitions, namely $\hbar(\om_\mathrm{u}-\om_\mathrm{l})=2\times 30~\mu\mathrm{eV}=0.06~$meV, where the phonon density is small, and $\hbar(\om_\mathrm{u}-\om_\mathrm{l})=2\times 200~\mu\mathrm{eV}=0.4~$meV, where it is much larger.
The TCL only considers the long-time limit of the dephasing rate in Fig. \ref{fig:multi_fig_g_explain}(a), explaining why the indistinguishability tends to unity for small QD-cavity coupling strengths.

To verify the intuitive explanation provided by the polariton picture, we show in Figs. \ref{fig:multi_fig_g_explain}(c) and (d) the phonon distribution function, $\langle\ti{b}^\dagger_{p}(t)\ti b_{p}(t)\rangle$, calculated using the ED approach.
For small QD-cavity coupling
no specific phonon energy is singled out, consistent with the small phonon density at the corresponding energy of $0.06~$meV, whereas for the larger QD-cavity coupling a significant increase in phonon population occurs near $0.4~$meV, as expected from the polariton interpretation.

The nearly constant indistinguishability for the ED and IBM for small $g$ arises due to the large difference in the involved timescales, the QD decay time, $\tau_\mrm{QD}=\Gamma_\mrm{eff}^{-1}$ with $\gamma=0$, and the extent of the short-time non-Markovian regime, $\tau_\mrm{corr}$ [Fig. \ref{fig:multi_fig_system_twotime_ID_g}(b)], where virtual processes dominate the decoherence \cite{Nazir2009b}.
The importance of the short-time regime is witnessed by the large dephasing rate in Fig. \ref{fig:multi_fig_g_explain}(a), which is especially important for the QD, and $\tau_\mrm{QD}$ needs to become comparable to $\tau_\mrm{corr}$ to affect the phonon dephasing.
We note that for $g \rightarrow 0$ other phonon dephasing mechanisms can become important \cite{Zimmermann2002,Muljarov2004}.
Further increasing $g$ in Figs. \ref{fig:multi_fig_system_twotime_ID_g}(c) and (d), real processes become increasingly important and contribute further to the decoherence, however, a stronger Purcell enhancement is also in effect, combating the influence of decoherence by making the QD decay faster.
Including only virtual phonon processes (IBM), the Purcell effect can increase the indistinguishability, however, adding real processes (ED) the indistinguishability is seen to decrease monotonically, partly due the saturation of the Purcell enhancement, and hence $\tau_\mrm{QD}$, in the strong coupling regime.
Comparing the indistinguishabilities for the QD and the cavity, only the ED predicts a significant difference between the two, which indicates that the difference arises from short-time non-Markovian effects that are only retained in the ED.
The smaller indistinguishability found for the QD is
a result of the direct interaction between the QD and the phonons, where the very strong short-time dephasing [Fig. \ref{fig:multi_fig_g_explain}(a)] significantly decreases the indistinguishability of photons emitted from the QD.
The photons in the cavity do not interact directly with the phonons, only indirectly through the QD-cavity interaction, and hence do not suffer to the same degree from the strong short-time dephasing as the QD does.
Furthermore, the longer lifetime of the QD compared to the cavity, i.e. $\tau_\mrm{QD}\ggg 1/\kappa$, is also expected to have an influence as excitations residing in the QD simply have more time to interact with the phonons.
We note that in the case of the phenomenological Lindblad theory, the same difference is not expected as here no short-time dephasing is present, only the constant pure dephasing rate $\gamma$.

\emph{Spectral asymmetries.} --
Figures \ref{fig:ID_vs_detun}(a) and (b) show the indistinguishability as a function of the QD-cavity detuning, $\Delta$, which is an important experimentally controllable parameter.
The Lindblad theory is unable to explain the variations with detuning that are predicted by the ED, both on a quantitive and qualitative level.
The behavior of the Lindblad theory can again be understood using the analytical expression discussed above, since the Purcell enhancement decreases for increasing detuning.
A common feature displayed by both the TCL and ED is a strong asymmetry with respect to the sign of the detuning.
For large detuning, $\abs\Delta\gg g$, the polariton dispersion becomes $\om_\mathrm{u/l}\approx(\Delta\pm\abs\Delta)/2$.
Thus, to make real transitions between the two polariton branches, the phonons need to provide an energy $\pm\hbar\abs\Delta$, either through emission $(-)$ or absorption $(+)$. 
At $T=0$, only positive detuning will lead to phonon emisson and thus decoherence in the long-time limit, hence a larger indistinguishability is expected for negative detuning.
Indeed, both the TCL and ED display such an asymmetry.
The effect of detuning on phonon emission is shown in Figs. \ref{fig:ID_vs_detun}(c), (d), and (e), where, for $\hbar\Delta=0.5~$meV, a significant phonon population is observed at this phonon energy.

While the detuning asymmetry of the cavity emission directly reflects the shape of the effective phonon density in Fig. \ref{fig:multi_fig_g_explain}(b), this is not the case for the QD emission.
The reason for this, perhaps surprising, difference between cavity and QD emission is the following: To generate a cavity photon, the QD must decay by coupling to the cavity, i.e. subject to the Purcell effect.
For large detuning ($\Delta \gg g,\kappa$), the Purcell effect is only effective if assisted by phonon emission \cite{Hohenester2009c,Kaer2010,Madsen}.
However, the QD can generate a photon without coupling to the cavity, namely through the background decay rate $\Gamma$.
Therefore we expect the cavity to significantly influence the QD emission only relatively close to resonance.
For large detunings, we expect the QD result to converge towards the $g\rightarrow 0$ result \cite{Note4}, predicted in Fig. \ref{fig:multi_fig_system_twotime_ID_g}(c).

For negative detuning, the effect of virtual processes in the short-time regime is clearly seen in the ED result, where despite the absence of phonon emission in the long-time limit, the indistinguishability is still significantly below unity, especially for QD emission. This is not the case for the TCL which only describes the long-time limit.

In conclusion, we have shown that non-Markovian phonon interactions strongly influence the coherence of single photons emitted from a cavity QED system.
An exact diagonalization approach predicts an upper limit for the indistinguishablity, a feature not captured by the commonly used Lindblad theory.
We provided physical insight into the
non-Markovian dephasing processes using a polariton picture.
Finally, we predict an asymmetry in the indistinguishability with respect to the QD-cavity detuning.

\textbf{Acknowledgments} -- The Center for Nanostructured Graphene (CNG) is sponsored by the Danish National Research Foundation, Project DNRF58. We thank Villum Fonden for financial support via the NATEC Center of Excellence.



%


\end{document}